\documentclass{article}

\PassOptionsToPackage{numbers, compress}{natbib}
\usepackage{listings}
\usepackage[frozencache,cachedir=.]{minted}

\usepackage[final]{nips_2018}

\usepackage[utf8]{inputenc} % allow utf-8 input
\usepackage[T1]{fontenc}    % use 8-bit T1 fonts
\usepackage{hyperref}       % hyperlinks
\usepackage{url}            % simple URL typesetting
\usepackage{booktabs}       % professional-quality tables
\usepackage{amsfonts}       % blackboard math symbols
\usepackage{nicefrac}       % compact symbols for 1/2, etc.
\usepackage{microtype}      % microtypography
\usepackage{graphicx}
\usepackage{xcolor}

\title{Just-in-Time Dynamic-Batching}
\author{
  Sheng Zha$^1$, Ziheng Jiang$^2$, Haibin Lin$^1$, Zhi Zhang$^1$\\
  $^1$Amazon AI, Amazon Web Services\\
  $^2$Paul G. Allen School, University of Washington\\
  \texttt{\{zhasheng,haibilin,zhiz\}@amazon.com}\\
  \texttt{ziheng@cs.washington.edu}
}

\begin{document}

\maketitle

\newcommand{\tf}{Tensorflow}
\newcommand{\tff}{Tensorflow Fold}
\newcommand{\dn}{DyNet}
\newcommand{\mb}{MatchBox}
\newcommand\hl[1]{\textcolor{orange}{[HL: #1]}}
\newcommand\sz[1]{\textcolor{red}{[SZ: #1]}}
\newcommand\hh[1]{\textcolor{blue}{[HH: #1]}}
\renewcommand{\theFancyVerbLine}{
  \sffamily\textcolor[rgb]{0.5,0.5,0.5}{\scriptsize\arabic{FancyVerbLine}}}

\begin{abstract}
Batching is an essential technique to improve computation efficiency in deep learning frameworks.
%Unlike in static-graph models, batching in models with dynamic computation graphs is highly non-trivial.
While batch processing for models with static feed-forward computation graphs is straightforward to implement, batching for dynamic computation graphs such as syntax trees or social network graphs is challenging due to variable computation graph structure across samples.
%Such deep learning models are required for modeling variable-structure data such as those found in languages and naturally occurring graphs.
%The need for efficient processing of these dynamic graph models give rise to automatic dynamic batching techniques in deep learning frameworks, such as those found in Tensorflow Fold and automatic-batching in Dynet.
%We identify the limitations of existing dynamic batching methods in on-line and off-line processing scenarios.
%Through simulation and analysis on a common workload of Tree-LSTM with variable number of child nodes, we pinpoint where performance can be further improved.
%wasted batching opportunity
%graph transformation overhead
Through simulation and analysis of a Tree-LSTM model, we show the key trade-off between graph analysis time and batching effectiveness in dynamic batching.
Based on this finding, we propose a dynamic batching method as an extension to MXNet Gluon's just-in-time compilation (JIT) framework.
We show empirically that our method yields up to 6.25 times speed-up on a common dynamic workload, a tree-LSTM model for the semantic relatedness task.
% We show empirically that our method yields significant speed-up on common dynamic workloads, such as Google Neural Machine Translation, tree-LSTM for semantic relatedness, and faster RCNN for object detection, all with just one line of code change.
% We further discuss more just-in-time compilation techniques that are made possible in our framework and furture directions.
% Existing methods such as \tff{} ~\cite{looks:fold} and \dn{} ~\cite{neubig:dybatch} uses heuristic design choices for balancing the two,
% which still have large improvement space according to our analysis.
% To close the gap, we propose new methods to adaptively learn the decisions by reinforcement learning.
%Our preliminary analysis show that the proposed methods reduce the size of the graph transformation problem by 33.75x in on-line processing, and increase the batch-able population by 60\% (TODO) in off-line processing.
\end{abstract}

\maketitle
\section{Introduction}
As modern neural networks evolve, dynamic neural network structures such as variable-length sequences~\cite{DBLP:journals/corr/abs-1810-04805, peters2018deep}, trees~\cite{tai:tree, tai2015improved}, and graphs~\cite{kipf2016semi, dai2018learning} become increasingly important.
For instance, social networks have complex graph structures, and natural language processing (NLP) problems often have variable-length sequences, each accompanied by a different parse tree~\cite{tai:tree}.
While most deep learning frameworks (e.g. Tensorflow~\cite{abadi2016tensorflow}, PyTorch~\cite{paszke2017automatic}, dyNet~\cite{neubig:dynet}, and Apache MXNet~\cite{chen:mxnet}) can express these dynamic neural networks as computation graphs, training with dynamic graphs can be much slower compared to static ones.
This is because for each sample, the framework needs to construct different computation graphs, move data across the memory hierarchy, which limits parallelism and efficient utilization of hardware resources.
Also, optimization techniques such as memory planning might not be effectively reused due to the variations in the computation graphs.
%This is because executing dynamic graph with a single instance suffers from the overhead of graph construction, data movement between memory hierarchies and limit parallelism, which results in low utilization of hardware resources.
Batch processing amortizes these overhead, provides better data locality, and enjoys better parallelism.
For example, many vector-vector multiplication operations that share the first operand can be batched into one matrix-vector multiplication.

% In computer vision models, the static model architectures and fixed input size make it easy to perform batching through mini-batches of examples.
% However, for problems with example-dependent input structures, batching is not straightforward.
% For instance, social networks have complex graph structures, and natural language processing (NLP) problems often have variable-length sequences, each having different parse trees.
%Advanced neural network models such as Tree-LSTM CITE can different RNNs depending on the parse tree of a sentence.

%However, applying batching to dynamic neural networks is non-trivial.
%Even the simple cases \hh{such as?} require careful preprocessing from users and incurs additional computation overhead.

However, naive batching technique is often accompanied by additional data preprocessing steps such as padding and bucketing for mini-batching variable-length sentences.
Models with dynamic graph structures often require more complex data preprocessing, such as ingesting parse tree data according for a tree-based model.
%Such techniques, however, are far from sufficient for
In this paper, we focus on dynamic batching methods that can automatically rewrite computation graphs to enable batching without extra data preprocessing, in a just-in-time fashion.

% Batching is a common technique in deep learning frameworks to improve computation efficiency.
% Through batching, samples sharing the same computation graph can be processed together, which results in better memory reuse and benefits from caching in the memory hierarchy in modern computers.
% Advanced neural network models for natural language tasks
% often result in varying computation graphs for each example.
% For instance, Tree-LSTM CITE builds a different RNN depending on the parse tree of a sentence.
%However, unlike static-graph models, batching for models with dynamic computation graphs is non-trivial.
% Such problems require careful preprocessing from users and incur additional computation overhead,
% such as padding and bucketing for variable-length sentences.
%For models with even more variations in the graph structures such as Tree-LSTMs and graph-LSTMs, simple batching techniques no longer work.
%On the other hand, the support for dynamic graphs is essential to modeling data with naturally occurring variable graph structures, such as those from languages and graphs.
% This work focuses on \emph{dynamic batching} methods that automatically rewrite computation graphs to perform batching without user preprocessing.
%such as Tensorflow Fold and Dynet automatic batching.
\begin{figure}
\centering
\includegraphics[height=1in, width=2.3in]{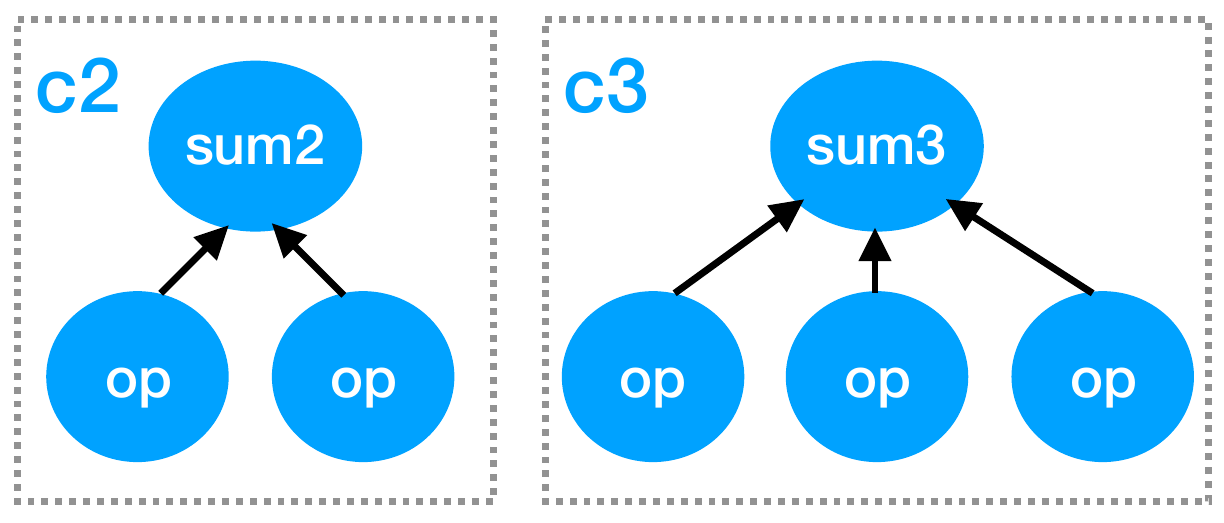}
\caption{Example computation graphs for Tree-LSTM. The outer frame marks the subgraph (tree) boundary. Each node represents an operator, and each $op$ node represents the same operation with the same parameterization. When analyzing these graphs at operator level, all leaf nodes can be batched. However, when analyzed at the subgraph level, C2 and C3 cannot be batched because they are not isomorphic.}
\label{fig:tree}
\end{figure}

%Two popular dynamic batching methods are implemented in \tff{} and \dn{} CITE?.
%Automatic dynamic batching requires additional graph analysis which has overhead.
% Dynamic batching methods usually consists of the following steps:
% \begin{enumerate}
% \item Construct the computation graph given user specification.
% \item Inspect the graph to identify batchable structures that need the exact same operations.
% \item Rewrite the graph to combine batched structures into one computation node.
% \end{enumerate}
Given a computation graph, the goal of dynamic batching is to identify subgraphs that can be executed together.
These subgraphs must be isomorphic in the sense of same graph topology, same node types, and same parameterization.
The isomorphism check guarantees consistent results between the original computation graph and our proposed batching method.
Each subgraph must also be capable of performing data-parallel (i.e. single instruction, multiple threads, SIMT) computation for a stack of samples to enable rewriting them into a single batched subgraph.

An unfortunate fact is that the analysis to identify isomorphic nodes at fine-grain require significantly more time than coarse-grain analysis.
For example, consider the computation graph in Figure~\ref{fig:tree}.
In this case, all leaf nodes are of the same type and can be batched together. Batching all leaf nodes results in the maximum computation reduction, but the isomorphism check is required on all 7 nodes.
On the other hand, if we take the perspective of a lower granularity and consider the two subgraphs C2 and C3, we only need to compare the two subgraphs with sum nodes as roots.
This reduces the analysis overhead but C2 and C3 are no longer co-exist in same batch.
Therefore, selecting the right granularity is vital to the efficiency of dynamic batching algorithms.

In this paper, we experimentally analyze the trade-off between graph processing time and individual subgraph batching speed-up. Through simulation we show that  on a Tree-LSTM~\cite{tai:tree} model, it is important to utilize the available granularities in user code and its effectiveness is obvious.
Based on this insight, we present just-in-time batching, a dynamic batching method well-suited for the just-in-time compilation in Apache MXNet Gluon~\cite{chen:mxnet}.
Our experiments indicate that our method performs significantly better than traditional batching methods by fully utilizing the subgraph structures under guidance in user code, with only one line of code change.

% Related work
\section{Related Work}
Some previous work approaches the batching problem from an application-specific perspective, such as cellular batching for RNNs with variable length inputs~\cite{gao2018low}, and batched graph execution~\cite{dgl}.
There also exist several general purpose dynamic batching methods, including \tff{}~\cite{looks:fold}, \dn{}~\cite{neubig:dybatch}, \mb{}~\cite{bradbury:matchbox}, and Cavs~\cite{zhang2017cavs}.
In this work, we focus on the general purpose methods.

Tensorflow Fold rewrites the computation graph before execution by matching graph node depth, and converts dynamic graphs into graphs with control-flow operations.
This approach is widely applicable but limited in two ways. Due to the granularity chosen at the time of graph constructions (i.e. subgraph level batching), this approach cannot batch operators that exist at finer granularity (e.g. addition and matmul).
As a result, some subgraphs cannot be batched even if they only vary in minor ways, such as trees with variable number of children (Figure~\ref{fig:tree}). Fold would treat such variants as completely independent components in the graph and produce inefficient rewritten graphs.
Furthermore, because the graph transformation happens before execution, this approach is less applicable when workload appears incrementally at irregular cadence while previous load is still being executed.
Such workload is commonly seen in model serving, where model inference requests can appear at any time.
By performing dynamic batching as part of JIT, our approach can handle such cases with good batching efficiency.

\dn{} auto-batching approach overcomes the above limitations by analyzing at the operator level online during execution.
The auto-batching analysis is undertaken in the scheduler. Kernels with the same signatures within the frontier of the running operators will be batched if they satisfy input dependencies.
In order to make sure that there are enough nodes in the pool to construct a potentially large batch, this approach depends on a heuristic to decide whether to wait for more nodes to arrive at the frontier before executing.
For workloads that involve many kernels, the analysis overhead can become a bottleneck and cannot be hidden through asynchronous execution, and thus dominates the overall processing time.
In contrast, benefit from various granularities available in MXNet Gluon, our approach enjoys freedom to choose a lower granularity to reduce the run-time analysis overhead.

\mb{} takes a similar approach to \tff{} but in the context of imperative execution in PyTorch.
At the time when \mb{} was devised, PyTorch executes the operators in a blocking way.
Thus, \mb{} creates stubs for all functional calls to collect the AST for batching analysis.
During execution, the computation graphs are dynamically constructed at runtime.
Our approach can instead happen as part of the Gluon JIT and cache the rewriting of graphs.

Cavs~\cite{zhang2017cavs} proposes a vertex-centric programming interface that decompose a dynamic neural network with two components: a static vertex function that can be batched together, and a sample specific graph which is dynamic. Our approach, built on top of MXNet Gluon's JIT, does not force models to be programmed as vertex and graph functions, and only requires one line of code change to enable batching.

\section{Effects of Subgraph Granularity on Dynamic Batching}
\label{sec:analysis}
To demonstrate the significance of the choice on granularity, we use a Tree-LSTM model in a simulation and compare the effectiveness and analysis overhead from two different choices of granularity: kernel level and subgraph (cell) level.
In the simulation, we apply the Fold method to batch 256 samples at a time, and record the effective batching-ratio, as well as the zoom-ratio by varying the granularity.
With same workload, batching ratio is defined as the ratio of the number of kernel launches with and without batching techniques.
The larger batching ratios achieved, more operators or subgraphs are being executed together, therefore higher speed-up can be achieved.

As shown in Table ~\ref{table:sickmetric}, we observe orders of magnitude difference between no-batch counts of kernels and subgraph, respectively. The difference between the batch ratios (1930x vs 137x) of kernels and subgraphs is strong evidence that the benefit of batching at fine-grain can be significant.
Therefore, it is vital to pick a granularity to balance reduced the analysis overhead and potential batching speed-up benefits.

% \begin{itemize}
%   \item Batch-able ratio: the ratio of the number of nodes without batching, to the smallest possible number of nodes in the batched graph. This can be estimated from the graph structures in the workload along with analysis on the model structure, if known.
%   \item Effective batching-ratio: the average ratio of the number of nodes without batching, to the number of nodes in the batched graph.
%   \item Zoom-ratio: on a given set of workload, the average ratio of the number of nodes in a graph at operator (fine) granularity, to the number of nodes at subgraph (coarse) granularity.
% \end{itemize}
% We assume that the reduction in number of nodes is proportional to reduction in analysis time.

\begin{table}
\centering
\label{table:sickmetric}
\caption{Statistics of kernels and subgraphs for Tree-LSTM on Sentences Involving Compositional Knowledge (SICK) dataset. The nodes in the trees from SICK dataset have varying number of children between 0 and 9. \textbf{No-batch}: the kernel launch count if no batching happens. \textbf{Batch}: the kernel launch count if batching happens. \textbf{Ratio}: batching ratio. The batched count for kernels is calculated from input graphs in the dataset. Count for subgraph batching is observed through simulation.}
\begin{tabular}{@{}l|lll@{}}
\toprule
Nodes     & kernel               & subgraph \\ \midrule
No-batch  & 5018658              & 148681    \\ \midrule
Batch     & \textasciitilde 2650 & 1081      \\ \midrule
Ratio     & 1930x                & 137x      \\ \bottomrule
\end{tabular}
\end{table}

% We implement the Tensorflow-Fold approach and observe the effect of dynamic batching by examining the number of batched subgraphs and the resulting reduction in number of nodes from converting the fine-grained operator-level graph.

% We simulate the batching effects on a real dynamic-graph workload using subgraph level Fold-based batching.
% The simulation is performed on child-sum tree-LSTM with variable number of children, for the semantic-relatedness task on Sentences Involving Compositional Knowledge (SICK) dataset.
% The parse trees in this dataset consist of nodes with 0 to 9 children.
In order to understand such difference, we closely examine the Tree-LSTM structure.
Each LSTM cell variant consists of 33 operators, where only 4 operators would vary based on the number of children. The rest of the operators have the exact same graph structure.
Due to the fact that cells with different number of children cannot be batched, the batching capabilities of the rest operators are ruined by these four dynamic operators. 
% If the granularity is chosen at cell level, due to the variance in these 4 operators, cells with different number of children cannot be batched, even though the rest of the operators can all benefit from batching.

% the large gap between batch-able ratio and effective batching-ratio. Besides, we can also observe the zoom-ratio of 33.75 from dividing on the no-batch row, which reflects the large potential saving in graph analysis complexity in \dn{} auto-batching in on-line scenario.

% The batchable ratio for operator-level graph constitutes a theoretical upper-bound for improvement from any batching method. Decreasing the granularity increases the subgraph size, as well as the side-effect of aforementioned subgraph mismatch problem that causes missed batching opportunity. On the other hand, decreasing the granularity can at the same time reduce the complexity for batching analysis by reducing the number of nodes.

\section{Just-in-time Dynamic Batching}

\begin{figure}[h]
\centering
\includegraphics[width=0.95\linewidth]{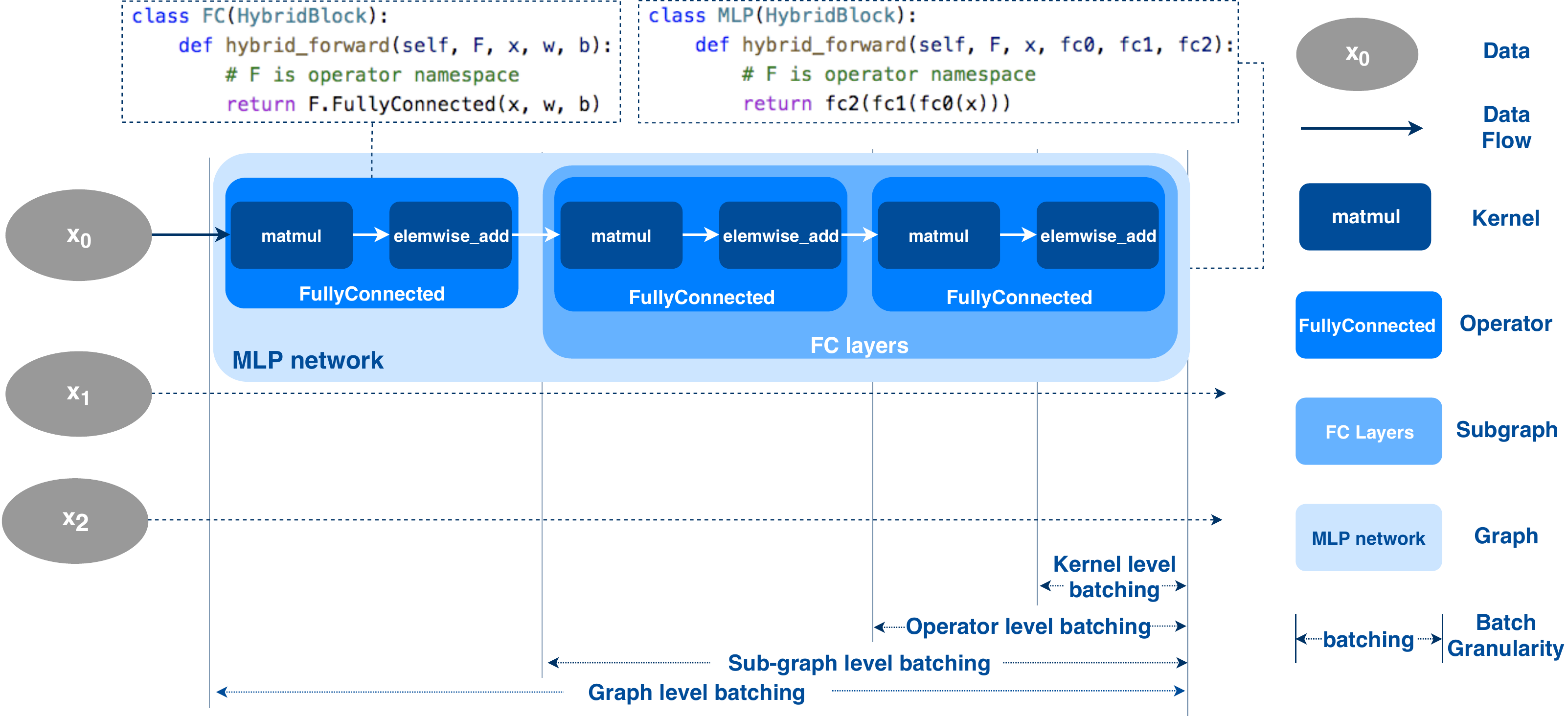}
\caption{Illustration of batching granularity in a multi-layer perceptron network. Individual data is passed through stacked fully connected layers (activation layers are omitted for simplicity). Gluon supports multiple granularity separation levels. Aside from graph-level batching which is traditional batching mechanism, subgraph, operator, and kernel level batching provide more granularity potentials for optimization.}
\label{fig:granularity}
\end{figure}

There are several factors to consider in the design of the dynamic batching methods.
From the analysis on Tree-LSTM model (Section~\ref{sec:analysis}), we recognize that the key to high performance batching scheme is choosing the right granularity.
Such granularity are often readily available from user code, because programmers are reasonably good at organizing code in a modular way to reduce repetition.
Gluon HybridBlock supports user-defined subgraphs at various levels, therefore we can take advantage of it to decide batching granularity during analysis (Section~\ref{sec:granularity}).
Our batching method also supports imperative execution in Gluon, which is integrated into Gluon's JIT~\ref{ref:eager}.

\subsection{Analysis Granularity}
\label{sec:granularity}
We visit the dynamic batching problem with MXNet's Gluon interface as it provides great potentials in analyzing and fulfilling different levels of batching granularity.
The Gluon interface offers a useful high-level abstraction, called "HybridBlock", which can be used as building block for the repeated computation subgraph for models with dynamic computation graphs, such as RNN cells.
Through just-in-time compilation (JIT), it creates and caches the recurrent cell logic into a symbolic graph, which is passed to execution engine as an operator node with a subgraph inside.

The subgraph is a loosely tied operator union (e.g. two layers of consecutive fully-connected layers) that can be cache-and-batched as a whole processing unit.
The operators again usually consist of series of instructions that can be further divided into finer-grained computations.
If needed, a fully connected operator can be further divided into two kernels: matrix multiplication and element-wise addition.
Such flexibility not only enables the versatile needs of expressing models, but also allows us to pick the level of granularity we need.

\subsection{Lazy Execution for Imperative Mode}
\label{ref:eager}
In addition, Gluon supports imperative execution, on top of which JIT converts to symbolic execution mode, and optimize and cache the computation graph.
Thus, our batching method should also support both deferred imperative execution, and operate in symbolic mode.
Supporting dynamic-batching in the symbolic programming style is straightforward, because it is possible to examine the complete computation graphs for all samples.
However, this does not hold for supporting imperative execution.

The NDArray interface in MXNet supports eager execution, in which we can only examine one small part of the computation graph at a time.
%Thus we need to first extend this abstraction to support dynamic batching.
We extend NDArray to support lazy execution through a new interface called NDArrayFuture.
Unlike regular NDArray which schedules the computation as soon as requests arrive, this interface allows delaying computation until user code requests to collect any of the resulting arrays.
This interface benefits from the great usability from regular imperative programming, such as easy debugging since users can request for the values of any array at anytime.
By delaying computation, it also sees more complete computation graphs, thus lending to more opportunity to optimize.

The operator registration mechanism in MXNet allows us to easily get an overview of all supported operators, and generate stub code according these operators, so that these NDArrayFuture instances can behave exactly like regular NDArrays but can be evaluated lazily.
Each time a new NDArrayFuture reference is created through the generated interface, we save the corresponding computation and organize the nodes and the input arguments in a look-up table according to their depth.
The nodes at the same depth are independent of each other and thus can be evaluated in parallel.
In order to identify the nodes that can be batched together, we use the computation node type, the node settings, the input argument layouts, as well as result look-up index to form a unique look-up key.

Further, we devise a new dynamic batching scope to allow users to easily specify the computation and the number of samples that we should try to batch.
When entering such scope, the interface automatically create the aforementioned look-up table, whose references are kept in the NDArrayFuture.
Within the scope, the execution of the computation of all the NDArrayFuture is delayed, until the code reaches the exit of the scope.
When exiting the scope, the computation is started, and is evaluated in the order of the computation graph depth.
The execution results are actual NDArrays, and are filled in the look-up table.
After exiting the batching scope, the NDArrayFuture can refer to the actual NDArray results and behave like regular NDArrays.

\subsection{Dynamic Batching}

Once we collected the graphs, we reorganize them into a look-up table so that the computation nodes that can be batched together reside in the same slot.
When executing, we stack these samples on the batch axis, and feed the newly formed batch to the common subgraph and launch execution.
After execution, we slice the output NDArray to obtain the results that correspond to individual samples.
Since such graph-rewrite can be expressed with symbolic execution, and MXNet supports control flows as operators, the graph rewriting can be cached and stored for next forward pass.
This also means that through delayed execution, we make dynamic batching as part of the JIT optimization in MXNet.
Last but not least, MXNet currently expresses control-flow also with operators, which means they are also nodes in the computation graphs.
Control-flow operations such as for-loop are often used to express repetition, exposing its inner details to the analysis logic is also important to achieve the best performance.

In our implementation, the user only need to add one extra line of code to declare the batching scope to enable dynamic batching, as shown in the following pseudo python code block.

% \begin{lstlisting}
% // network and loss definition
% net = GraphConvolutionNet()
% loss = SoftmaxCELoss()
% // batching scope
% with mx.batching():
%     // iterate through samples in the batch
%     for data, label in data_batch:
%         // forward computation
%         out = net(data)
%         ls = loss(out, label)
%         // backward computation
%         ls.backward()
% // parameter update
% trainer.step()
% \end{lstlisting}

\begin{minted}[mathescape,
               linenos,
               numbersep=5pt,
               gobble=2,
               frame=lines,
               framesep=2mm]{python}
  #network and loss definition
  net = GraphConvolutionNet()
  loss = SoftmaxCELoss()
  #batching scope
  with mx.batching():
      #iterate through samples in the batch
      for data, label in data_batch:
          #forward computation
          out = net(data)
          ls = loss(out, label)
          #backward computation
          ls.backward()
  #parameter update
  trainer.step()
\end{minted}

\section{Experiments}

% In this section, we evaluate JIT dynamic batching with three different workloads: Tree-LSTM, Faster-RCNN and Google Neural Machine Translation.

% \subsection{Tree-LSTM}

In this section, we benchmark the training and inference speed of tree-structured LSTM networks on the semantic-relatedness task on SICK dataset~\cite{marelli:sick} for both training and inference.
SICK dataset consists of 4500 pairs of sentences, each labeled with a semantic-relatedness score for the two sentences.
We use Stanford Parser~\cite{manning:corenlp} for extracting the parse trees of the sentences.
The nodes in the parse trees from SICK dataset have varying number of children between 0 and 9.

The benchmark runs on EC2 c4.8xlarge instance with Intel Xeon E5-2666 v3 (Haswell) processors.
The per-instance method evaluates one input instance at a time, while JIT dynamic batching uses a batch size of 256.
With dynamic batching we are able to achieve 5.96x speedup for training and 6.25x speedup for inference.

\begin{table}[h]
\centering
\caption{Training and inference speed of Tree LSTM on SICK dataset.}
\label{table:sickmetric}
\begin{tabular}{@{}l|lll@{}}
\toprule
Method     & Training (samples/s)  & Inference (samples/s) \\ \midrule
Per instance  & 33.77               & 50.46    \\ \midrule
JIT dynamic-batching     & 201.11 (5.96x) & 315.54 (6.25x)      \\ \midrule
\end{tabular}
\end{table}

% \subsection{Faster-RCNN}

% In order to test different workloads, we also report benchmark results using Faster-RCNN \cite{ren2015faster} which is one of the most popular object detection model.
% It is also known as the foundation of other state-of-the-art object detection models.
% Faster-RCNN consists of two stage network: region proposal network and pooling based regressor networks, respectively.
% Due to the properties of dynamic input shapes and multiple staging execution path, there is no naive batching strategy other than padding the input data.
% Therefore Faster-RCNN networks have great speed up potentials using our proposed dynamic batching.

% \begin{table}
% \centering
% \caption{Inference speed comparison of Faster-RCNN on COCO 2017 dataset.}
% \label{table:frcnnmetric}
% \begin{tabular}{@{}l|ll@{}}
% \toprule
% Method       & Inference (samples/s) \\ \midrule
% Per instance                 &     \\ \midrule
% JIT dynamic-batching      &  (x)      \\ \midrule
% \end{tabular}
% \end{table}

% \subsection{Google Neural Machine Translation}

% The Google Neural Machine Translation (GNMT) model is an deep LSTM-based network with 8 encoder and 8 decoder layers with with attention and residual connections~\cite{wu:gnmt}.
% The network takes a variable length source language sequence as the input to the encoder, and uses beam search during decoding to find the target variable length sequence.

\section{Conclusion}

% contributions again
% limitations and future directions
Deep learning models with dynamic computation graphs are increasingly important in the research community.
%Because of the difficulty in manually batching the computation in these models, automatic dynamic batching techniques are important in facilitating the research and application that require modeling dynamic structures in data.
In this paper, we identify the key design choice for dynamic batching algorithms, the granularity of the subgraph for analysis, which affects both analysis time and discover-ability of computation available for batching.
Based on the design of JIT in MXNet, we design a batching method that is suitable for both imperative execution and symbolic execution under JIT.
Our method utilizes the existing hierarchy in user code, which provides good starting point in terms of subgraph granularity.
Our experiments show the effectiveness of this simplistic approach.

Furthermore, the asynchronous extension of NDArray, i.e. our new NDArrayFuture interface, provides an extensible basis for other computation optimization such as operator fusion, function approximation. Its flexibility will facilitate the prototyping of optimization techniques in future work.
% Our proposed learning-based method can improve the computation performance by learning to trade-off analysis time and batching effectiveness.
% We plan to implement and evaluate the proposed method and experiment on a diverse set of real-life workloads.
%Another interesting direction to explore is to use recent advances in algorithms for subgraph isomorphism for the purpose of extracting batchable parts.

\bibliographystyle{plainnat}
\bibliography{dynamic-batching}

\end{document}